\documentclass[conference]{IEEEtran}

\usepackage{amsmath}
\usepackage{amssymb}
\usepackage{graphicx}
\usepackage{subfigure}
\usepackage{cite}

\newtheorem{theo}{Theorem}
\newtheorem{prop}{Proposition}

\IEEEoverridecommandlockouts
\begin{document}

\title{QoS Analysis of Cognitive Radios\\Employing HARQ}

\author{Sami Akin, Marwan Hammouda, and J\"{u}rgen Peissig\\
Institute of Communications Technology\\
Leibniz Universit\"{a}t Hannover\\
Email: \{sami.akin, marwan.hammouda, and peissig\}@ikt.uni-hannover.de
\thanks{This work was supported by the European Research Council under Starting Grant--306644.}}

\maketitle
\begin{abstract}
Recently, the demand for faster and more reliable data transmission has brought up complex communications systems. As a result, it has become more difficult to carry out closed-form solutions that can provide insight about performance levels. In this paper, different from the existing research, we study a cognitive radio system that employs hybrid-automatic-repeat-request (HARQ) protocols under quality-of-service (QoS) constraints. We assume that the secondary users access the spectrum by utilizing a strategy that is a combination of underlay and interweave access techniques. Considering that the secondary users imperfectly perform channel sensing in order to detect the active primary users and that there is a transmission deadline for each data packet at the secondary transmitter buffer, we formulate the state-transition model of the system. Then, we obtain the state-transition probabilities when HARQ-chase combining is adopted. Subsequently, we provide the packet-loss rate in the channel and achieve the effective capacity. Finally, we substantiate our analytical derivations with numerical results.
\end{abstract}

\section{Introduction}
Recently, communications systems evolve into more complex systems due to increasing demand for faster and more reliable data transmission. Herein, cognitive radios are one of the systems that are proposed as future's data transmission technologies \cite{mitola2000cognitive}. The purpose behind cognitive radio systems is to provide a solution to the inefficient use of the spectrum by opening the spectrum to the secondary (unlicensed) users along with the primary (licensed) users \cite{haykin2005cognitive}. In cognitive radio systems, three well-known spectrum sharing methods for the secondary users are overlay, underlay and interweave spectrum access techniques \cite{ghosh2014cognitive}. The secondary users can operate simultaneously with the primary users in the overlay and underlay techniques as long as they act in accordance with certain constraints, while they have to avoid transmission when the primary users are active in the interweave technique. In all these techniques, the secondary users should be aware of the activities of the primary users. Therefore, one required functionality in cognitive radio systems is spectrum sensing, i.e., detecting the unused spectrum and sharing the spectrum without harmful interference with other users \cite{akyildiz2006next}. However, detecting the unused spectrum is not always perfect. In particular, there are miss-detections and false alarms \cite{duan2010cooperative}.

Meanwhile, flawless data transmission is another objective in wireless communications. However, due to the dynamic nature of wireless media, transmission errors occur easily. Especially when the channel fading is too strong, it becomes difficult to sustain data transmission at a certain rate with a minimized bit or packet error probability. Therefore, error detection and correction techniques such as automatic-repeat-request (ARQ) and forward error correction are employed as certain protocols during the transmission of data. Additionally, a hybrid of these two, which is called hybrid-ARQ (HARQ), was proposed in order to provide better error correction performance \cite{wozencraft1961coding,lin1982hybrid,mandelbaum1974adaptive}. Following the introduction of HARQ, several HARQ protocols were suggested. Among these protocols, HARQ chase combining (HARQ-CC) became one of the widely-known method of packet combining technique \cite{chase1985code}. The repeated transmissions of a packet are linearly added at the receiver in order to obtain a power gain in HARQ-CC.

Because HARQ systems may require the retransmission of transmitted packets or send additional data packets due to decoding failures at receivers, delay-sensitive traffic concerns have become a research focus as well. From this perspective, HARQ systems were analyzed in \cite{gunaseelan2010performance,villa2012adaptive,anastasopoulos2008delay}. Moreover, employing effective capacity as the performance metric, which is the maximum sustainable data arrival rate at a transmitter buffer that a stochastic service process in a channel can support under quality-of-service (QoS) constraints, the authors in \cite{li2015throughput,li2014energy} obtained a closed-form effective capacity expression. However, the closed-form expression is valid under less strict QoS constraints. Concurrently, the authors in \cite{akin2015backlog} provide the effective capacity expression for a general class of HARQ systems in point-to-point transmission scenarios.

In this paper, we investigate the performance of cognitive radio systems that employ HARQ protocols under QoS constraints by taking effective capacity into consideration. We assume that the secondary users access the spectrum by engaging a combined strategy of the underlay and interweave access techniques. To the best of our knowledge, the effective capacity performance of cognitive radios that utilize HARQ protocols have not been researched yet. The rest of the paper is organized as follows: We provide the cognitive radio system model in Section \ref{sec:cognitive_radio}. In particular, we describe the channel sensing technique, the channel input-output relation and the HARQ protocol. Furthermore, we analyze the system performance in Section \ref{sec:performance_analysis} by attaining the state-transition model, the packet-loss probability and the effective capacity.

\section{Cognitive Radio System}\label{sec:cognitive_radio}
\begin{figure}
\begin{center}
\includegraphics[scale=0.35]{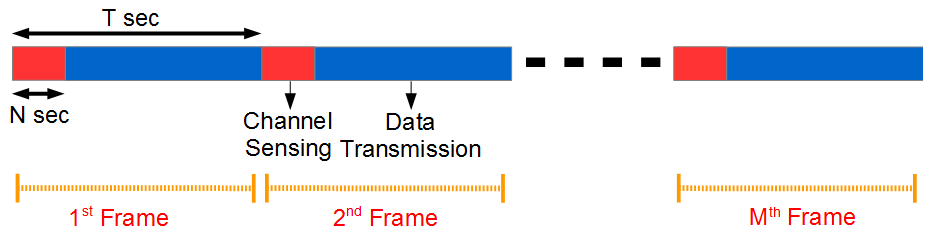}
\caption{Frame structure of cognitive radio transmission.}\label{res:res_1}
\end{center}
\end{figure}

We focus on a point-to-point transmission scenario in which one secondary transmitter and one secondary receiver communicate with each other in the presence of primary users. We assume that the secondary users perform communication in two phases: channel sensing and data transmission. They initially sense a transmission channel in the first $N$ seconds of a frame duration $T$ seconds in order to detect the active primary users in that channel. Then, they transmit data in the remaining $T-N$ seconds. This communication framework is displayed in Figure \ref{res:res_1}. We further assume that the primary users are active in one transmission frame with probability $\rho$ and that their activity status changes from one frame to another independently. Following the channel sensing phase, the secondary transmitter selects the transmission power policy in the data transmission phase in order to limit the interference inflicted on the active primary receivers. Particularly, the average symbol power is set to $P_{b}$ when the channel sensing result admits a busy state, whereas the average symbol power is $P_{i}$ when the channel sensing result is idle. In general, the following setting is considered: $P_{b}\leq P_{i}$. At the same time, the available bandwidth in each transmission channel is $B$ Hz.

\subsection{Channel Sensing}
If the transmission strategies of the primary users are not known, energy-based detection methods are well-established in channel sensing processes. Therefore, we adapt the energy detection technique to our transmission model. We also note that other detection techniques can be easily adapted to the transmission scenario and the problem formulation. Because our focus is on HARQ in cognitive radios in this paper, we employ energy-based detection methods for brevity and a straightforward analysis. Hence, noting that the channel sensing in the above model is conducted for $N$ seconds in a channel with a bandwidth of $B$ Hz, we have $NB$ complex symbols during the sensing phase. Then, we express mathematically the hypothesis testing problem between the noise and the signal in noise in each frame as follows:
\begin{alignat}{3}
\mathcal{H}_{i}&:\quad&&y(t)=w(t) &&\quad t=1,\cdots,NB,\\
\mathcal{H}_{b}&:\quad&&y(t)=w(t)+s(t) &&\quad t=1,\cdots,NB,
\end{alignat}
where $w(t)$ is the noise at the secondary receiver, $s(t)$ is the primary transmitters' faded signal arriving at the secondary receiver\footnote{We consider a channel sensing process employed at the secondary receiver. However, the aforementioned hypothesis testing can be performed either at the secondary receiver or at the secondary transmitter or at any other sensing device or collaboratively. If the sensing mechanism is incorporated at a device other than the secondary receiver, we need to consider the noise and signal variables at that device rather than $w(t)$ and $s(t)$, respectively.}, and $y(t)$ is the complex channel output at the secondary receiver. Above, $\{w(t)\}$ forms an independent and identically distributed (i.i.d.) sequence of additive, zero-mean, circularly symmetric, complex Gaussian random noise variables with variance $\mathbb{E}\{|w(t)|^2\}=\sigma_{w}^{2}<\infty$. Likewise, $s(t)$ is also an i.i.d. zero-mean, circularly symmetric, complex Gaussian random variable with variance $\mathbb{E}\{|s(t)|^2\}=\sigma_{s}^{2}<\infty$. This is a valid assumption in certain channel scenarios \cite{akin2010effective}. Now, invoking the Neyman-Pearson detector for the above testing problem \cite{poor1994introduction}, we have
\begin{equation}
Y=\frac{1}{NB}\sum_{t=1}^{NB}|y(t)|^{2}\gtrless^{\mathcal{H}_{b}}_{\mathcal{H}_{i}}\lambda,
\end{equation}
where $\lambda$ is the detection threshold. Noting that $Y$ is chi-square distributed with $2NB$ degrees of freedom, we establish the probabilities of false alarm and detection as follows:
\begin{align}
p_{f}&=\Pr\{\mathcal{\widehat{H}}_{b}|\mathcal{H}_{i}\}=\Pr\{Y>\lambda|\mathcal{H}_{i}\}\nonumber\\
&=1-p\left(\frac{NB\lambda}{\sigma_{w}^{2}},NB\right),\label{false_alarm}\\
p_{d}&=\Pr\{\mathcal{\widehat{H}}_{b}|\mathcal{H}_{b}\}=\Pr\{Y>\lambda|\mathcal{H}_{b}\}\nonumber\\
&=1-p\left(\frac{NB\lambda}{\sigma_{w}^{2}+\sigma_{s}^{2}},NB\right),\label{detection}
\end{align}
respectively. Above, $p(x,a)$ is the regularized lower incomplete Gamma function and is defined as $p(x,a)=\frac{1}{\Gamma(a)}\int_{0}^{x}\tau^{a-1}e^{-\tau}d\tau$ where $\Gamma(a)$ is the Gamma function. $\mathcal{\widehat{H}}_{b}$ and $\mathcal{\widehat{H}}_{i}$ denote the busy and idle detection results, respectively. The false alarm probability is the probability of having a busy sensing result given that the channel is actually free of the primary users. The detection probability is the probability of detecting the active primary users given that the channel is actually occupied by the primary users. Basically, the miss-detection probability is $1-p_{d}$.

\begin{figure*}
\begin{center}
\includegraphics[scale=0.43]{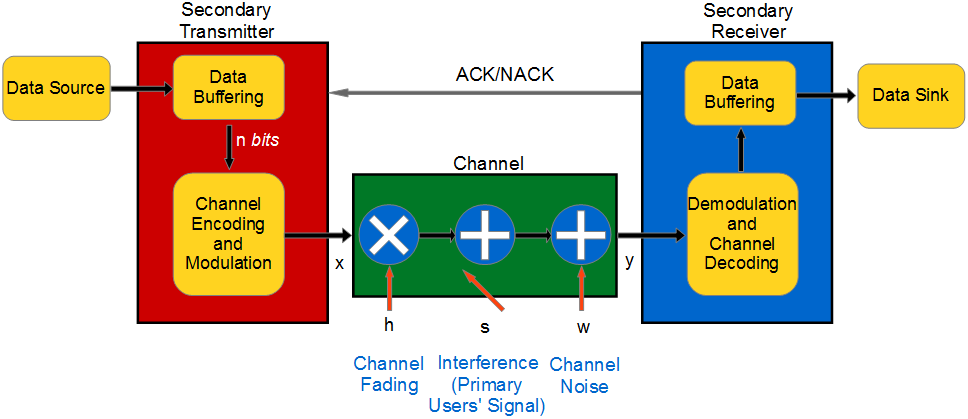}
\caption{Channel model. The transmitter initially stores the data packets in its buffer, subsequently encodes, modulates and forwards each data packet to the receiver through the wireless channel. Then, the receiver feeds the transmitter with ACK/NACK depending on its decoding performance. The channel fading coefficient, $h$, and the thermal noise, $w$, are present in the channel continuously, while the signal emitted by the primary users, $s$, appears intermittently.}\label{res:res_2}
\end{center}
\end{figure*}
\subsection{Input-Output Relation During Data Transmission}
The discrete-time input-output relation during the transmission of the data in one frame in the channel is given as
\begin{equation}\label{input_output_idle}
y(t)=hx(t)+w(t)\quad t=NB+1,\cdots,TB,
\end{equation}
when the primary users are idle, and
\begin{equation}\label{input_output_busy}
y(t)=hx(t)+s(t)+w(t)\quad t=NB+1,\cdots,TB,
\end{equation}
when the primary users are active. Above, $x(t)$ is the complex channel input at the secondary transmitter. The channel fading between the secondary transmitter and the secondary receiver is denoted by $h$, and it has an arbitrary distribution with finite average power $\mathbb{E}\left\{|h|^2\right\}=\mathbb{E}\left\{z\right\}=\sigma_{h}^{2}<\infty$. Here and throughout the paper, $z = |h|^2$ represents the magnitude-square of the fading coefficient. We consider a block-fading channel, i.e., $h$ is constant during one transmission frame and changes independently from one frame to another. We further assume that the channel side information is available at the secondary receiver, i.e., the secondary receiver knows the actual value of $h$, while the secondary transmitter is aware of the channel statistics only. Moreover, as a result of a limited power budget, the channel input is subject to the following average power constraints: $\mathbb{E}\{|x(t)|^2\}\leq P_{b}/B$ and $\mathbb{E}\{|x(t)|^2\}\leq P_{i}/B$ when the channel is sensed as busy and idle, respectively. Since we assume that $B$ complex symbols per second are transmitted, the average power of the system is constrained by $P_{b}$ and $P_{i}$ when the channel sensing result is busy and idle, respectively.

\subsection{HARQ}
As seen in Fig. \ref{res:res_2},  we assume that the secondary transmitter initially divides the data received from the source into packets of $n$ bits and stores them in its buffer. Subsequently, it implements the encoding, modulation, and transmission of each packet in frames of $T$ seconds with the first-come first-served policy. Since the secondary users spend $N$ seconds for channel sensing, the data transmission phase indeed takes $T-N$ seconds in one transmission frame. During the transmission of a packet, if the packet is decoded correctly by the secondary receiver in any time frame, the secondary receiver sends a positive acknowledgment (ACK) to the secondary transmitter, and the packet is removed from the transmitter buffer. Otherwise, the secondary receiver sends a negative ACK (NACK), and the secondary transmitter retransmits the same encoded and modulated packet, or it sends additional information to alleviate the decoding process at the secondary receiver. At the receiver side, if a packet is received with errors in one time frame, the secondary receiver may either discard the erroneously received packet or keep it to utilize it in the next time frame. Furthermore, each packet should be transmitted in $MT$ seconds due to a transmission deadline where $M\in\{1,2,\cdots,\}$. In detail, if a packet is not received correctly at the end of the $M^{\text{th}}$ transmission trial, it will be discarded from the transmitter queue, regardless of decoding performance at the secondary receiver. Therefore, we consider a removal of any packet from the transmitter buffer, either as a result of successful decoding by the secondary receiver or due to the transmission deadline, as a packet service from the queue.

\section{Performance Analysis}\label{sec:performance_analysis}
Regarding the decision of channel sensing and the activities of the primary users, we have four possible channel scenarios:
\begin{enumerate}
\item Channel is busy, and detected as busy (correct detection) with probability $q_1=\rho p_{d}$,
\item Channel is busy, but detected as idle (miss-detection) with probability $q_2=\rho(1-p_{d})$,
\item Channel is idle, but detected as busy (false alarm) with probability $q_3=(1-\rho)p_{f}$,
\item Channel is idle, and detected as idle (correct detection) with probability $q_4=(1-\rho)(1-p_{f})$.
\end{enumerate}

Recall that the average transmission power is $P_{b}$ when the channel is sensed as busy and the average transmission power is $P_{i}$ when the channel is sensed as idle. Hence, we express the instantaneous transmission capacity in each scenario as follows: $C_{\text{i}}=B\log_{2}(1+\zeta_{\text{i}}z_{\text{l}})$ where $\zeta_{\text{i}}$ is the signal-to-noise ratio in the corresponding scenario. Particularly, $\zeta_{1}=\frac{P_{b}}{\sigma_{w}^{2}+\sigma_{s}^{2}}$, $\zeta_{2}=\frac{P_{i}}{\sigma_{w}^{2}+\sigma_{s}^{2}}$, $\zeta_{3}=\frac{P_{b}}{\sigma_{w}^{2}}$ and $\zeta_{4}=\frac{P_{i}}{\sigma_{w}^{2}}$. We denote the channel fading power in the $l^{\text{th}}$ time frame by $z_{l}$. The instantaneous capacities are achievable when $x(t)$ is complex, zero-mean Gaussian distributed, i.e., $x(t)\sim\mathcal{CN}(0,P_b/B)$ when the channel sensing result is busy and $x(t)\sim\mathcal{CN}(0,P_i/B)$ when the channel sensing result is idle \cite[Ch. 9.1]{book_information_theory}. However this requires the availability of channel side information, $h_{l}$, at the secondary transmitter. Furthermore, due to the channel sensing errors, the secondary users do not know exactly which scenario they are in. Because $h_{l}$ is most likely not available at the secondary transmitter in practice and there are possibly channel sensing errors, we consider a more practical scenario, where a wireless link is equipped with an HARQ protocol, which can be modeled as a discrete-time finite state Markov chain.

\subsection{State-Transition Model}\label{sec:state_transition_model}
\begin{figure}
\begin{center}
\includegraphics[scale=0.32]{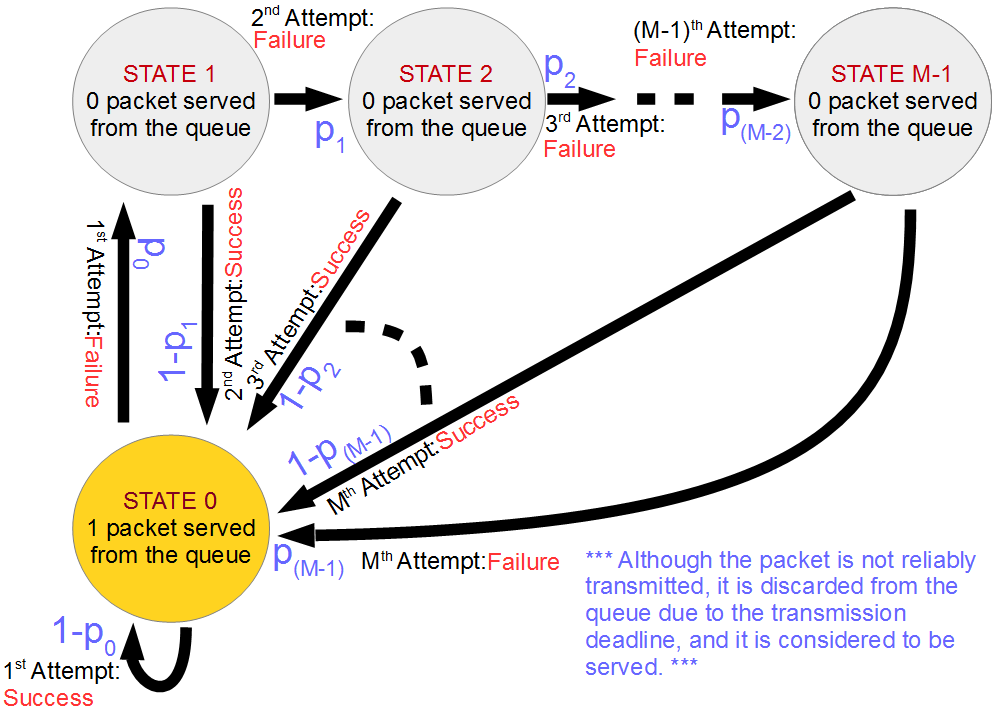}
\caption{State transition model. The system enters State 0 when a packet is removed from the secondary transmitter buffer. The other states represent the decoding failures following the data transmission attempts.}\label{res:res_3}
\end{center}
\end{figure}
For an analytical presentation, we model the queue activity at the end of each time frame as a discrete-time Markov process. We assume that there are $M$ states. While State 0 comes after any packet removal from the transmitter buffer at the end of any time frame, the other states represent the decoding failures following the transmission attempts. As seen in Fig. \ref{res:res_3}, when a packet is removed (served) from the transmitter buffer following a successful decoding at the end of the $m^{\text{th}}$ transmission attempt of the packet, the system enters State 0 with probability $1-p_{m-1}$ where $m\in\{1,\cdots,M-1\}$. In other words, the system being in State $m-1$ goes to State 0 when there is a success in decoding the packet at the end of the $m^{\text{th}}$ transmission attempt. Furthermore, when a decoding failure occurs, the system goes to State $m$ with probability $p_{m-1}$. On the other hand, the system enters State 0 at the end of the $M^{\text{th}}$ transmission attempt with probability 1 regardless of the decoding performance at the secondary receiver, because the transmitted packet is removed from the queue due to the transmission deadline. Note that a packet transmitted for the $M^{\text{th}}$ time is decoded either correctly with probability $1-p_{M-1}$ or incorrectly with probability $p_{M-1}$. Correspondingly, we can provide the following state transition matrix $\Phi$:
\begin{equation}\label{phi}
\Phi=\begin{pmatrix}
    1-p_0 & 1-p_1 & 1-p_2 & \cdots & 1-p_{M-2} & 1\\
    p_0 & 0 & 0 & \cdots & 0 & 0\\
    0 & p_1 & 0 & \cdots & 0 & 0\\
    0 & 0 & p_2 & \cdots & 0 & 0\\
    \vdots &\vdots & \vdots& \ddots&\vdots&\vdots\\
    0 & 0 & 0 & \cdots & p_{M-2} & 0
  \end{pmatrix}.\nonumber
\end{equation}

In the following proposition, we provide the state transition probabilities when HARQ-CC is employed by the secondary users. In the HARQ-CC protocol, if a packet is received with errors by the secondary receiver in any frame, the secondary receiver keeps the incorrectly received packet and the secondary transmitter retransmits the same encoded and modulated packet. With every retransmission, the secondary receiver implements maximum-ratio-combining until the packet is decoded correctly unless the transmission deadline is reached at the secondary transmitter. For more details, we refer to \cite{chase1985code}.
\begin{prop}\label{prop:state_transition_probability_cognitive_radio}
Given that the secondary users employ the HARQ-CC protocol and that a packet of $n$ bits is encoded and modulated into $(T-N)B$ symbols in one time frame, the state-transition from State 0 to State 1 is given as
\begin{align}\label{state_transition_pro_0}
p_{0}&=1-\sum_{i=1}^{4}q_{i}e^{-\frac{\kappa}{\zeta_{l}^{i}\sigma_{h}^{2}}},
\end{align}
where $\kappa=2^\frac{n}{(T-N)B}-1$. Above, $q_{1}$, $q_{2}$, $q_{3}$ and $q_{4}$ are the probabilities of the aforementioned channel scenarios described in Section \ref{sec:performance_analysis}, and $\zeta_{l}^{1}$, $\zeta_{l}^{2}$, $\zeta_{l}^{3}$ and $\zeta_{l}^{4}$ are the signal-to-noise ratios in the corresponding scenarios. Likewise, the state transition from State $m$ to State $m+1$ is given as
\begingroup
\allowdisplaybreaks
\begin{align}\label{professor}
\hspace{0.0cm}p_{m}=&\sum_{m_1,m_2,m_3,m_4}^{m}\frac{m!\prod_{i=1}^{4}q_i^{m_i}}{F(\kappa,m_1,m_2,m_3,m_4)\prod_{i=1}^{4}m_i!}\nonumber\\
&\times\Big\{q_1F(\kappa,m_1+1,m_2,m_3,m_4)\nonumber\\
&\hspace{0.5cm}+q_2F(\kappa,m_1,m_2+1,m_3,m_4)\nonumber\\
&\hspace{0.5cm}+q_3F(\kappa,m_1,m_2,m_3+1,m_4)\nonumber\\
&\hspace{0.5cm}+q_4F(\kappa,m_1,m_2,m_3,m_4+1)\Big\}
\end{align}
\endgroup
for $m\in\{1,\cdots,M-1\}$, where $m_{i}$ indicates the number of transmission frames in which we have Scenario $i$ in the last $m$ transmission attempts or time frames, i.e., $m=\sum_{i=1}^{4}m_i$. Above,
\begin{equation*}
F(\kappa,a,b,c,d)=C\sum_{k=0}^{\infty}\delta_{k}p(\kappa,a+b+c+d+k),
\end{equation*}
$C=\zeta_{\min}\prod_{i=1}^{4}(\zeta^{i})^{-m_i}$, $\zeta^{\min}=\min\{\zeta^1\cdots,\zeta^4\}$, $\delta_{0}=1$, $\delta_{k}=\frac{1}{k+1}\sum_{l=0}^{k+1}l\gamma_{l}\delta_{k+1-l}$ and $\gamma_{k}=\sum_{i=1}^{4}\frac{m_i}{k}(1-\zeta^{\min}/\zeta^{i})^{k}$. Recall that $p(x,a)$ is the regularized lower incomplete Gamma function and is defined in (\ref{false_alarm}) and (\ref{detection}).
\end{prop}
\emph{Proof:} Omitted due to space limitation.$\hfill$$\square$

Note that the state-transition probability from State $M-1$ to State 0 is 1 and that $p_{M-1}$ in (\ref{professor}) provides the probability of decoding failure given that the system is in State $M-1$.

\subsection{Packet-Loss Probability}
Now, let $\pi=[\pi_{0},\cdots,\pi_{M-1}]$ be the steady-state probability vector of the above state transition model given in Fig. \ref{res:res_3}, where $\sum_{m=0}^{M-1}\pi_{m}=1$ and $\pi=\Phi\pi$. Subsequently, we express the steady-state probabilities as follows:
\begin{equation}
\pi_{m}=\pi_{0}\prod_{j=0}^{m-1}p_{j}\text{ for }m\in\{1,\cdots,M-1\}
\end{equation}
and
\begin{equation}
\pi_{0}=\frac{1}{1+\sum_{m=1}^{M-1}\prod_{j=0}^{m-1}p_{j}}.
\end{equation}
Recall that at the end of the $M^{\text{th}}$ frame, the packets are removed from the buffer either as a result of decoding success or due to the transmission deadline. The packets that are removed due to the transmission deadline are considered as lost since they do not reach the data sink at the secondary receiver side. Therefore, we define the packet-loss probability as the ratio of the lost packets to the total number of packets that are removed (served) by the secondary transmitter, which is expressed as
\begin{equation}\label{dekanat}
p_{\text{lost}}=\frac{p_{M-1}\pi_{M-1}}{\pi_{0}}=p_{M-1}\prod_{m=0}^{M-2}p_{m}=\prod_{m=0}^{M-1}p_{m}.
\end{equation}
Above, $\pi_{0}$ is the steady-state probability of clearing the queue off one packet in one frame, while $p_{\text{lost}}$ is the probability of a packet not reaching the secondary receiver.

\subsection{Effective Capacity}
Notice in (\ref{dekanat}) that increasing the value of $M$ decreases the packet-loss probability. On the other hand, the average queueing delay of a packet in the transmitter buffer increases with increasing $M$. As a result, we need to accommodate another tool that takes the queueing delay and buffering measures into account. Therefore, we propose effective capacity that outlines the asymptotic decay rate of the buffer occupancy. Effective capacity establishes the maximum constant arrival rate that a given service process can support in order to guarantee desired statistical QoS constraints specified by the QoS exponent $\theta$ \cite{wu_negi}. Defining $Q(t)$ as the stationary queue length at the transmitter buffer at time $t$, i.e., the number of packets at the secondary transmitter buffer, and $\theta$ as the decay rate of the tail distribution of the queue length $Q(t)$, we can express the following:
\begin{equation*}
\lim_{q\to\infty}\frac{\log\Pr\{Q(t)\geq q\}}{q}=-\theta,
\end{equation*}
where $q$ is the buffer threshold. Therefore, we have the following approximation for larger $q$: $\Pr\{Q(t)\geq q\}\approx \text{e}^{-\theta q}$. So, larger $\theta$ refers to stricter buffer constraints, and smaller $\theta$ refers to looser buffer constraints. Furthermore, it is shown in \cite{liu_chamberland} that $\Pr\{D(t)\geq d\}\leq c\sqrt{\Pr\{Q(t)\geq q\}}$ for constant data arrival rates, where $D(t)$ is the steady-state delay of a packet in the buffer, and $c$ is a positive constant. Noting that $q=ad$, effective capacity provides us with the maximum arrival rate when the system is subject to the statistical queue length or delay constraints in the form of $\Pr\{Q(t)\geq q\}\leq\text{e}^{-\theta q}$ or $\Pr\{D(t)\geq d\}\leq c\text{e}^{-\theta ad/2}$, respectively. For a given QoS exponent $\theta>0$, effective capacity, $\rho_{S}(\theta)$, is given by
\begin{equation}\label{for_slack_term}
\rho_{S}(\theta)=-\frac{\Lambda(-\theta)}{\theta}=-\lim_{t\to\infty}\frac{1}{\theta t}\log_{e}E\{e^{-\theta S(0,t)}\}
\end{equation}
where $S(\tau,t)=\sum_{k=\tau+1}^{t}r(k)$ is the time-accumulated service process, and $r(k)$ for $k=1,2,...$ is the discrete-time, stationary and ergodic stochastic service process. We note that $\Lambda(\theta)$ is the asymptotic log-moment generating function of $S(0,t)$, and is given by $\Lambda(\theta)=\lim_{t\to\infty}\frac{1}{t}\log E\left\{e^{\theta S(0,t)}\right\}$. Henceforth in the next result, we provide the effective capacity of a cognitive radio system that employs HARQ protocols.
\begin{theo}\label{theo:effective_capacity}
For the aforementioned cognitive radio system that employs any HARQ protocol with the state-transition model given in Section \ref{sec:state_transition_model}, the effective capacity for a given QoS exponent $\theta$ is given by
\begin{equation}\label{effective_capacity_rho}
\rho_{S}(\theta)=-\frac{1}{\theta T}\log_{e}\left(\chi^{\star}\right)\text{ bits/sec,}
\end{equation}
where $\chi^{\star}$ is the only unique real positive root of $f(\chi)$, where
\begingroup
\allowdisplaybreaks
\begin{align}\label{f_yy}
\begin{split}
f&(\chi)=\chi^{M}-(1-p_{0})\text{e}^{-\theta n}\chi^{M-1}\\
-&\sum_{m=1}^{M-2}(1-p_{m})\chi^{M-1-m}\text{e}^{-\theta n}\prod_{j=0}^{m-1}p_{j}-\text{e}^{-\theta n}\prod_{j=0}^{M-1}p_{j}.
\end{split}
\end{align}
\endgroup
\end{theo}
\emph{Proof:} Omitted due to space limitation.$\hfill$$\square$

The real positive root of $f(\chi)$ can be found by using numerical techniques. For instance, bisection method can be efficiently used to find the solution since it has only one real positive root. Moreover, the aforementioned theorem is valid for cognitive radios employing any HARQ protocol. It is also worth mentioning that the effective capacity expression (\ref{effective_capacity_rho}) confirms the effective capacity of point-to-point transmission systems that employ HARQ protocols, which is given in \cite{akin2015backlog}.

\section{Numerical Results}
\begin{figure}
\begin{center}
\includegraphics[scale=0.3]{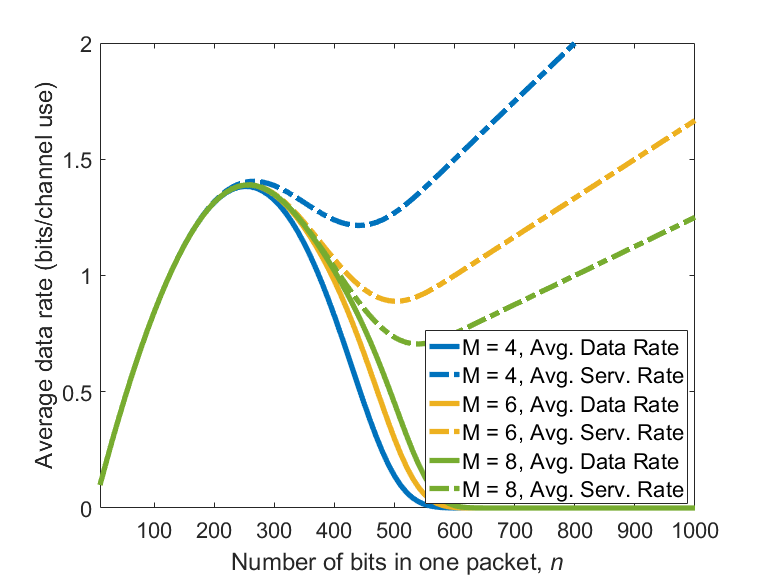}
\caption{Average transmission rate and service rate versus the number of bits in one transmission packet with different transmission deadlines. The solid lines indicate the average transmission rate arriving at the secondary transmitter and the dashed lines indicate the average data service rate in the channel.}\label{fig:fig_1}
\end{center}
\end{figure}
In this section, we present the numerical results to substantiate our analytical derivations. We assume that the aforementioned cognitive radio system is equipped with HARQ-CC. Unless otherwise indicated, we consider the following normalized settings for the sake of simplicity: $\sigma_{w}^{2}=1$, $\sigma_{s}^{2}=1$ and $\sigma_{h}^{2}=1$. We further assume that the channel fading is modeled with Rayleigh distribution while the above analysis is valid for any fading distribution with finite variance. We set the transmission frame duration to $10^{-4}$ seconds, the sensing duration to $2\times10^{-5}$ seconds and the channel bandwidth to $10^{6}$ Hz, i.e., $T=10^{-4}$, $N=2\times10^{-5}$ and $B=10^{6}$, respectively. Moreover, when the channel is detected as busy, the transmission power is set to $0$ dB, i.e., $10\log_{10}\frac{P_{b}}{B\sigma_{n}^{2}}=0$ dB, and when the channel is detected as idle, the transmission power is set to $10$ dB, i.e., $10\log_{10}\frac{P_{i}}{B\sigma_{n}^{2}}=10$ dB. Regarding the channel sensing settings, we have the energy detection threshold $\lambda=1.4$, where the number of channel samples is $NB=20$.

In Fig. \ref{fig:fig_1}, we plot the average data transmission rate to the secondary receiver, $\frac{\pi_0(1-p_{\text{lost}})n}{TB}$, and the average service (data removal) rate from the secondary transmitter buffer, $\frac{\pi_{0}n}{TB}$, as a function of the number of bits in one data packet. Herein, the difference between the average service rate from the buffer and the average transmission rate to the secondary receiver reveals the data-loss (packet-loss) rate in the channel. The solid lines show the rates at which data arrives at the secondary receiver and the dashed lines show the rates at which the secondary transmitter buffer is cleared. We have results for different transmission deadline values, $M$. The gap between the average data rate to the secondary receiver and the average service rate from the buffer is almost zero for small values of packet size. However, the gap increases with increasing packet size. In order to prevent packet losses, the secondary transmitter should send data at lower rates. One more interesting result is that when the packet size is small, the choice of transmission deadline is not necessary.

\begin{figure}
\begin{center}
\includegraphics[scale=0.3]{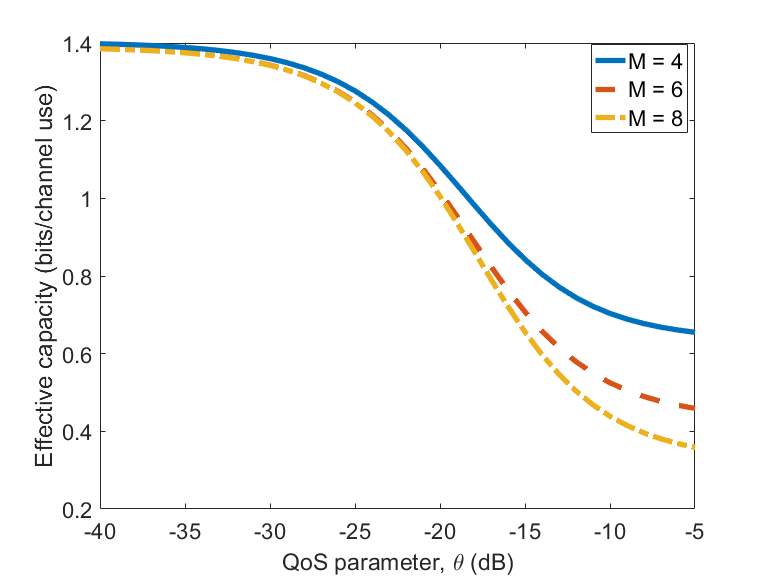}
\caption{Effective capacity versus queue decay rate parameter, $\theta$, with different transmission deadlines.}\label{fig:fig_2}
\end{center}
\end{figure}

\begin{figure}
\begin{center}
\includegraphics[scale=0.3]{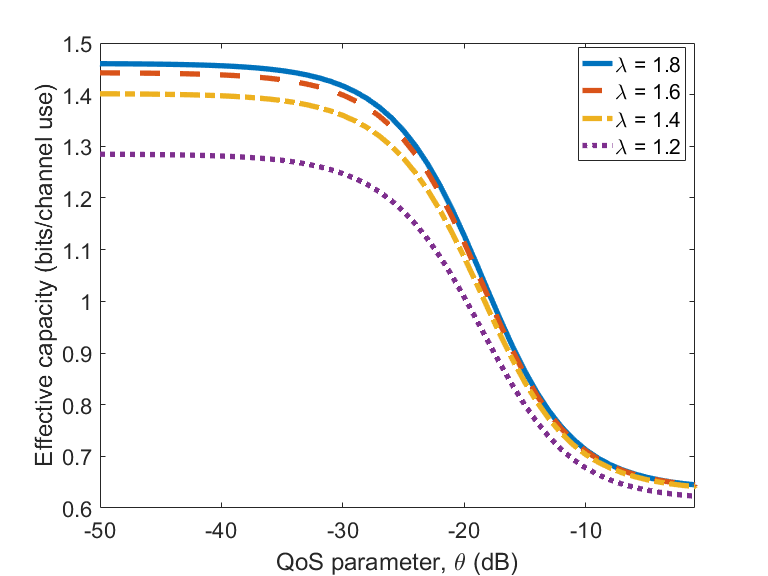}
\caption{Effective capacity versus queue decay rate parameter, $\theta$, with different energy sensing thresholds.}\label{fig:fig_3}
\end{center}
\end{figure}

In Fig. \ref{fig:fig_2}, we plot the effective capacity performance, i.e., the maximum sustainable data rate to the secondary transmitter buffer that the cognitive radio system employing HARQ-CC can support under defined QoS constraints, as a function of the QoS exponent (queue decay rate parameter) $\theta$ considering different transmission deadlines. In each plot, we set the number of bits in each packet equal to the optimal values obtained in Fig. \ref{fig:fig_1}. The optimal packet size is set to the value that maximizes the average data transmission rate to the secondary transmitter. Note that the data-loss rate is negligible. When $\theta$ is low, the effective capacity performance approaches the average data transmission rate. However, the effective capacity decreases with increasing $\theta$. Herein, the performance of the system when $M$ is small becomes better because when $\theta$ increases the effective capacity approaches the minimum data transmission rate to the secondary receiver, which is $\frac{n}{TBM}$ in bits per channel use. Likewise, we plot the effective capacity in Fig. \ref{fig:fig_3} with different energy detection threshold values when $M=4$. Increasing the detection threshold leads to more miss-detections, and hence, less protection for the primary users. Since, the secondary transmitter expends more energy when the channel is sensed as idle, the effective capacity performance increases. However, there is more interference on the primary receivers. The performance gap is higher when $\theta$ is small, i.e., under less strict QoS constraints. Nevertheless, the performance gap becomes insignificant with increasing $\theta$. Basically, when the QoS constraints is higher, the secondary user can sense the channel with a lower detection threshold and transmit data at lower rates so that the primary users are protected and the packet-loss rates are minimized.

\section{Conclusion}
In this paper, different from the existing studies, we have studied a cognitive radio system that employs HARQ protocols under QoS constraints. We have formulated state-transition model of the system and calculated the state-transition probabilities when HARQ-CC is applied. We have provided closed-form expression for the packet-loss probability in the channel and the effective capacity. We have shown that the choice of the transmission deadline is not necessary with small data packet sizes when regarding the average data transmission rate to the secondary receiver in the channel. We have also shown that when the QoS constraints are less strict, the choice of transmission deadline does not affect the system performance, whereas when the QoS constraints are much stricter, it is better to set a small value for the transmission deadline. We have further observed that the secondary users should avoid the miss-detection of the primary users as much as possible when the QoS constraints are stricter, because sending data at high power level does not result in an increase in the effective capacity performance.

\bibliographystyle{IEEEtran}
\bibliography{IEEEabrv,references}

\end{document}